\documentclass[a4paper,11pt]{article}
\pdfoutput=1 

\usepackage{jheppub1} 

\usepackage[latin3]{inputenc}
\usepackage[makeroom]{cancel}
\usepackage{graphicx}
\usepackage{amsmath}
\usepackage{amsfonts}
\usepackage{amssymb}
\usepackage{color}
\usepackage[export]{adjustbox}
\usepackage{graphicx}
\usepackage{dcolumn}
\usepackage{bm}
\usepackage{simplewick}
\usepackage{array}
\usepackage{appendix}
\usepackage{relsize}
\usepackage{subcaption}

\newcommand{\hR}{{}^*{R}}
\newcommand{\be}{\begin{equation}}
    \newcommand{\ee}{\end{equation}}
\newcommand{\beq}{\begin{equation}}
    \newcommand{\eeq}{\end{equation}}
\newcommand{\bea}{\begin{eqnarray}}
    \newcommand{\eea}{\end{eqnarray}}

\title{\boldmath Particle dynamics and geometric optics in Chern-Simons black holes}

\author[a,b]{Rehana Rahim}
\author[a]{and Khalid Saifullah}

\affiliation[a] {Department of Mathematics, Quaid-i-Azam University, Islamabad, Pakistan}
\affiliation [b]{Department of Mathematics and Statistics, Riphah International University, Islamabad, Pakistan}

\emailAdd{rehana.rahim@riphah.edu.pk}
\emailAdd{ksaifullah@fas.harvard.edu}
\abstract{In this paper we study the effects of the coupling constant of the Chern-Simons modified gravity
on some physical properties of black holes. The Hawking mass is one of
the proposed definitions of quasilocal mass. We find that, for slowly rotating
Chern-Simons black holes, the Hawking mass is independent of the
coupling constant. Next, we show the dependence on the centre of mass energy, for
two neutral colliding particles, of  coupling constant and the rotation parameter.
We also investigate energy extraction through Penrose process and find
that the energy gain and efficiency of the Penrose process are independent of
this coupling constant. Rotation of the polarization vector is also studied for dependence on the Chern-Simons coupling constant.
\vspace{80 mm}}

\notoc

\begin{document}
\maketitle
\flushbottom
\section{Introduction}
Extremely massive, rotating black holes are believed to be present at the
centre of most of galaxies. The gravitational field outside such black holes is
important not only for the evolution of the captured compact objects but is also
a source of gravitational waves. These gravitational waves have been
recently discovered \cite{1}. A qualitative description of such a
gravitational field is provided by the Kerr metric in the theory of general
relativity (GR). According to the no-hair theorem \cite{2a}, a black hole in general relativity, is described only by its mass, spin and electric charge.

The theory of GR has been tested extensively. One of the test is to see that the black holes are represented by the Kerr solution \cite{2b,2c,2d}.
Tests in the weak gravitational field regime can depend on the parametrized post-Newtonian (PPN) approach \cite{3}.
On the other hand, testing in the strong gravitational regime requires the construction of the
black hole metric in terms of parametric deviations from the Kerr spacetime \cite{4b,4c, 12,rs}. Such modifications in the theory of
gravity have received a lot of attention. This is mainly attributed to
quantum gravity which is requisite for the description of extreme situations
in the universe like the vicinity of a black hole.

One of the many interesting theories that modify gravity is the Chern-Simons (CS) theory
\cite{5,6}. The four-dimensional CS theory has some remarkable
features, e.g. (i) it can be obtained from the superstring theory where the CS
term is important for the cancellation of the anomaly in Lagrangian density
\cite{6,7}, (ii) the Schwarzschild solution remains valid in the CS
theory, hence all classical tests of GR hold in this theory
\cite{5}, and (iii) the  CS gravity theory modifies the axial part of the gravitational
field as compared to GR.

CS theory has two types of independent theoretical formulations, namely,
dynamical and non-dynamical. In the non-dynamical theory, the CS scalar is \textit{a priori} prescribed
function with the evolution equation becoming a differential constraint on the
space of allowed solutions. On the other hand, the dynamical theory has CS
scalar as a dynamical field having its own evolution equation and
stress-energy-momentum tensor. In recent years there has been an increasing interest in the dynamical CS theory.

The  black hole solutions have been developed in the non-dynamical theory \cite{8,9,10,11}.
A solution has been determined \cite{8,9} by employing far-field
approximation where $\varphi$ (the CS scalar field) is linearly proportional to
the asymptotic time coordinate $t$. This solution is stationary but not
axisymmetric and gives correction to the frame dragging effect. The slow rotation
approximation was employed to obtain a rotating black hole
solution \cite{10}. An exact rotating solution in non-dynamical theory was found \cite{11} which is stationary and axisymmetric for
arbitrary $\varphi$.
In dynamical theory, a solution in slow rotation approximation and small coupling constant was also obtained \cite{12,13}. This
solution was extended to include the terms for second order in spin parameter \cite{14}. The solution in slow rotation approximation up to the $5^{th}$ order in spin parameter has also been found \cite{14a}.
The assumption of slow rotation was relaxed in Ref. \cite{15} where CS scalar field induced by a (rapidly) rotating black hole (Kerr metric) in the dynamical theory was considered.

In this work, we investigate black holes under slow rotation approximation in
dynamical CS theory and study their different properties. Our focus is to see how
the coupling constant of CS theory effects different physical phenomena e.g.
quasi local mass, particle motion, energy extraction process and the geometrical optics. We will show
that the Hawking mass and efficiency of the Penrose process do not depend on the
coupling constant, whereas, it is interesting to note that the centre of mass energy and the polarization angle show a behaviour that is dependent on the coupling constant. All the computations are done to the second order in
the spin parameter and to the order $a\gamma^{2}$ in the coupling constant where  $a$ is the spin parameter and $\gamma$ is CS coupling constant.

The paper is organized as follows. In Section \ref{CHT} we will give a brief
review of mathematical formalism of the CS theory. After this introduction, the subsequent sections will focus on different physical phenomena. In Section \ref{HM} we
will discuss Hawking mass in CS theory and study its relation to the coupling
constant. Section \ref{ECM} provides an analytical expression for the
centre of mass energy for two colliding neutral particles as a function of CS
coupling constant.  Energy extraction through Penrose process is
discussed in Section \ref{PP}. Rotation of the polarization vector is the subject of Section \ref{rotpol}.
We conclude the work with a brief summary in the
last section. In this work, we take $G=c=1$ and the coordinates are taken in the order
$(t,r,\theta,\phi)$ with indices running from $0$ to $3$.
\section{The Chern-Simons theory}\label{CHT}
The action for a CS gravity
is written as \cite{null}
\begin{equation}
S=k\int\sqrt{-g}Rdx^{4}+\frac{\gamma}{4}\int dx^{4}\sqrt{-g}\varphi \hR R
 -\frac{1}{2}\int\sqrt{-g}(\nabla
\varphi)^{2}dx^{4},
\label{ac}
\end{equation}
where $\gamma$ is the CS coupling constant, $k=1/16\pi$, $\varphi$ is a scalar
field, $g$ denotes the metric tensor $g_{\mu\nu}$'s determinant ,
$R=g^{\lambda\alpha
}R_{\lambda\alpha}$ represents the Ricci scalar with $R_{\lambda\alpha}$ being the Ricci tensor, $\hR R$
is the Pontryagin density defined as
\begin{equation}
\hR R=\hR^{\mu\rho\sigma}_{\nu}R^{\nu}_{\mu\rho\sigma},
\end{equation}
where $\hR^{\mu\rho\sigma}_{\nu}=\varepsilon^{\rho\sigma\delta\gamma}R^{\mu}_{\nu\delta\gamma}/2 $ is the dual Riemann tensor.
Here $\varepsilon^{\rho\sigma\delta\gamma}$ represents the 4 dimensional Levi-Civita tensor.
The first term is the standard Einstein-Hilbert action, the second term
is the CS correction term and the third term is the scalar field term. This action is parity even.
Varying the action with respect to $g_{\mu\nu}$ and the scalar field
gives two equations, respectively%
\begin{equation}
R_{\mu\nu}-\frac{1}{2}Rg_{\mu\nu}+\frac{\gamma}{k}C_{\mu\nu}=\frac{1}%
{2k}\nabla_{\mu}\varphi\nabla_{\nu}\varphi-\frac{1}{4k}g_{\mu\nu}\nabla_{\rho
}\varphi\nabla^{\rho}\varphi, \label{2}
\end{equation}
\begin{equation}
\nabla_{\mu}\nabla^{\mu}\varphi+\frac{\gamma}{4}\ast R_{\mu\nu\rho\sigma
}R^{\nu\mu\rho\sigma}=0. \label{3}
\end{equation}
In Eq. (\ref{2}) $C_{\mu\nu}$ is the traceless
C-tensor \cite{11} given as
\begin{equation}
C^{\mu\nu}=\nabla_{\delta}\varphi\varepsilon^{\delta\rho\sigma\mu}%
\nabla_{\sigma}R_{\rho}^{\nu}+\nabla_{\delta}\nabla_{\rho}\varphi\ast
R^{\rho\mu\nu\delta}+\left[\mu\longleftrightarrow\nu\right].
\end{equation}
The exterior derivative of a CS form gives the Pontryagin density $\ast
R_{\mu\nu\rho\sigma}R^{\mu\nu\rho\sigma}$ as
\[
\ast R_{\mu\nu\rho\sigma}R^{\mu\nu\rho\sigma}=2\nabla_{\mu}\varepsilon
^{\mu\alpha\beta\lambda}\left[\Gamma_{\alpha\rho}^{\delta}\partial_{\beta}
\Gamma_{\lambda\delta}^{\rho}+\frac{2}{3}\Gamma_{\alpha\rho}^{\delta}
\Gamma_{\beta\sigma}^{\rho}\Gamma_{\lambda\delta}^{\sigma}\right],
\]
giving the conservation of the topological current. Using the above equation, the CS correction term in the action can be simplified by partial integration as
\begin{equation}
\frac{\gamma}{4}\int\sqrt{-g}\varphi\ast R_{\mu\nu
\rho\sigma}R^{\mu\nu\rho\sigma}dx^{4}=-\frac{\gamma}{2}\int dx^4\sqrt{-g}(\nabla_{\mu}\varphi)\varepsilon
^{\mu\alpha\beta\lambda}\left[\Gamma_{\alpha\rho}^{\delta}\partial_{\beta}
\Gamma_{\lambda\delta}^{\rho}+\frac{2}{3}\Gamma_{\alpha\rho}^{\delta}
\Gamma_{\beta\sigma}^{\rho}\Gamma_{\lambda\delta}^{\sigma}\right].
\end{equation}
The spinning solution of
Eqs. (\ref{2}) and (\ref{3}), valid for slow rotation
and small coupling constant, is given in Refs. \cite{12,13}. The full Kerr
spacetime is
\begin{equation}
\begin{split}
ds_{K}^{2}&=-\Big[1-\frac{2Mr}{\Sigma}\Big]dt^{2}-\frac{4aMr\sin^{2}\theta}{\Sigma
}dtd\phi+\frac{\Sigma}{\Delta}dr^{2}+\Sigma d\theta^{2}+\sin^{2}\theta
\Big[r^{2}+a^{2}
\nonumber
\\
&
+\frac{2a^{2}Mr\sin^{2}\theta}{\Sigma}\Big]d\phi^{2}, \label{4}
\end{split}
\end{equation}
where
\begin{equation}
\Sigma=r^{2}+a^{2}\cos^{2}\theta,\\
\quad
\Delta=r^{2}+a^{2}-2Mr.
\end{equation}
Here $M$ and $a$ denote mass and spin of the black hole respectively.
If we keep the terms up to $O(a^{2})$ in the slow rotation approximation
$a<<M,$ the Kerr metric takes the form
\begin{equation}
\begin{split}
ds_{SK}^{2} & =-\left[U+\frac{2a^{2}M\cos^{2}\theta}{r^{3}}\right]dt^{2}-\frac
{4aM\sin^{2}\theta}{r}dtd\phi+\frac{1}{U^{2}}
\left[U-\frac{a^{2}}{r^{2}}
\Big(1-U\cos^{2}\theta\Big)\right]dr^{2}
\nonumber
\\&
+\Sigma d\theta^{2}
+\sin^{2}\theta\Bigg[r^{2}+a^{2}
+\frac{2a^{2}M\sin^{2}\theta
}{r}\Bigg]d\phi^{2}, \label{7}
\end{split}
\end{equation}
where $U=1-\frac{2M}{r}$. The solution corresponding to the CS term is given
as \cite{12}
\begin{equation}
ds^{2}=ds_{SK}^{2}+\frac{5\gamma^{2}a\sin^{2}\theta}{4kr^{4}}\left[1+\frac{12M}
{7r}+\frac{27M^{2}}{10r^{2}}\right]dt d\phi. \label{8}
\end{equation}
The equation for the scalar field $\varphi$ is
\begin{equation}
\varphi=\left[\frac{5}{2}+\frac{5M}{r}+\frac{9M^{2}}{r^{2}}\right]\frac{\gamma
a\cos\theta}{4Mr^{2}}. \label{9}
\end{equation}
We note that the off-diagonal term which results in a weakened dragging effect has the coupling constant contribution to $O(a\gamma^{2})$. One can follow an outgoing quasispherical light cone backwards in time from
$ \mathcal{I}^{+}$ for the purpose of determining the event horizon. This null cone is given by the axisymmetric null hypersurface $v(t,r,\theta,\phi)=t-w(r,\theta)=constant$, which satisfies the equation
\begin{equation}
\partial_{\alpha} v \partial_{\beta }v  g^{\alpha\beta}=0.\end{equation}
On expanding this gives
\begin{equation}
g^{tt}+(\partial_{r} w)^{2} g^{rr}+(\partial_\theta w)^{2} g^{\theta\theta}=0.
\end{equation}
The inverse metric components appearing in the above equation are independent of CS correction, and thus lead to the same horizon as in the Kerr black hole \cite{12}.

\section{Hawking mass}\label{HM}

In GR gravitational field is a non-local object and pointwise
energy (or mass) cannot be defined for it. Although gravitational field mass
cannot be defined locally, it is still possible to define it on quasi local
level i.e. on a bounded region of spacetime. There are several definitions for
mass at quasi local level such as the Brown-York energy \cite{by}, the
Misner-Sharp mass \cite{sha}, the Komar mass \cite{ko}, the Bartnik mass
\cite{bar}, the Hawking mass \cite{haw}, the Geroch mass \cite{ger} and the
Penrose mass \cite{pen}. In this work, we will study Hawking mass because it is more convenient and appropriate for our purpose.
\par
Let $S$ be a spacelike 2-surface defined by constant $t$ and constant $r$. The area of the surface is denoted by $A$.
Consider a null tetrad $(l^{\mu},n^{\mu},m^{\mu}, \bar{m}^{\mu})$ on $S$. Here $l^{\mu}$ and $n^{\mu}$, respectively,
represent outgoing and ingoing future directed null vectors orthogonal to
$S$, and $m^{\mu},\bar{m}^{\mu}$ are the tangent vectors to $S$.
On $S,$ the Hawking mass is given by \cite{haw}
\begin{equation}
m_{H}(S)=\sqrt{\frac{A}{\left(  4\pi\right)^{3}}}\text{
}\left[2\pi+\int_{S} \rho\acute{\rho}dS\right], \label{10A}
\end{equation}
where $\rho$ and $\acute{\rho}$ denote the spin coefficients
of the Newman-Penrose formalism, representing the expansions of
the outgoing and ingoing null cones, respectively \cite{20}.

Hawking mass on the event horizons of Reissner-Nordstr\"om and Kerr black holes is
discussed in Ref. \cite{bq}. We discuss Hawking mass for spacetime
(\ref{8}) for the regions outside the event horizon.

The outgoing and ingoing null 4-vectors are given by \cite{12a}
\begin{align}
l^{\mu}&=\Bigg(\frac{r}{r-2M}-\frac{2Ma^{2}}{r\Big(r-2M\Big)^{2}},1,0,\frac{a}
{r\Big(r-2M\Big)} \nonumber-\frac{10a\pi\gamma^{2}}{r^{5}\Big(r-2M\Big)}\left[
1+\frac{12M}{7r}+\frac{27M^{2}}{10r^{2}}\right]\Bigg),
\end{align}
\begin{align}
n^{\mu}&=\Bigg(\frac{1}{2}+\frac{a^{2}\sin^{2}\theta}{2r^{2}},-\frac{\Big(r-2M\Big)}
{2r}+\frac{a^{2}\Big(r-2M\Big)\cos^{2}\theta}{2r^{3}}-\frac{a^{2}}{2r^{2}},0,\frac
{a}{2r^{2}}-\frac{5a\pi\gamma^{2}}{r^{6}}\Bigg[
1+\frac{12M}{7r}
\nonumber
\\&
+\frac{27M^{2}}{10r^{2}}\Bigg]\Bigg),
\nonumber
\end{align}
where we have multiplied the ingoing null 4-vector by $(r-2m)/{2r}+a^{2}/2r^{2}
-a^{2}(r-2M)\cos^{2}\theta/2r^{3}$ for the orthogonality condition
$\mathbf{l}\cdot\mathbf{n}=-1$ to hold and called it $n^{\mu}$. To determine complex null 4-vector, we
will employ certain properties that a null tetrad and a metric tensor satisfy
in Newman-Penrose formalism. Let us denote the complex null vector by
$\mathbf{m}$. In the component form it is represented as
\begin{equation}
m^{\mu}=(A,B,C,D), \label{25}
\end{equation}
and its complex conjugate as $\bar{m}^{\mu}=(\bar{A},\bar{B},\bar{C},\bar
{D})$. The null vectors $m^{\mu}$ and $\bar{m}^{\mu}$ satisfy the condition $\textbf{m}.\bar{\textbf{m}}=1$. In terms of  $(l^{\mu},n^{\mu},m^{\mu},\bar{m}^{\mu})$, the inverse metric tensor $g^{\mu\nu}$ is \cite{20}
\begin{equation}
g^{\mu\nu}=-l^{\mu}n^{\nu}-l^{\nu}n^{\mu}+m^{\mu}\bar{m}^{\nu}+m^{\nu}\bar
{m}^{\mu}. \label{27}
\end{equation}
By employing this and the orthogonality conditions $m_{\mu}m^{\mu}=0=l_{\mu}m^{\mu
}$, one can determine the
components of the null vector $m^{\mu}$ and its complex conjugate $\bar
{m}^{\mu}$. The expression of $m^{\mu}$  is given as
\begin{align}
m^{\mu}&=\Bigg( \frac{ia\sin\theta}{\sqrt{2}r}+\frac{a^{2}\cos\theta\sin\theta}
{\sqrt{2}r^{2}},0,\frac{1}{\sqrt{2}}\Bigg[ \frac{1}{r}-\frac{ia\cos\theta}{r^{2}}-\frac{a^{2}\cos^{2}\theta}{r^{3}}\Bigg] ,\frac{i}{\sqrt{2}r\sin\theta}+\frac
{a\cos\theta}{\sqrt{2}r^{2}\sin\theta}
\nonumber
\\&
-\frac{ia^{2}\cos^{2}\theta}{\sqrt
{2}r^{3}\sin\theta}\Bigg).
\label{46}
\end{align}
By replacing $i$ by $-i$, we can obtain $\bar m^{\mu}$.

There are $12$ spin coefficients in the Newman-Penrose formalism. For the sake of Hawking mass we require only two, namely, $\rho$ and $\rho^{\prime}$. In terms of the null vectors ($l^{\mu},n^{\mu},m^{\mu},\bar{m}^{\mu})$, their expressions are \cite{dem}
\begin{align}
&\rho=l_{\nu;\mu}m^{\nu}\bar{m}^{\mu},
\\&
\rho^{\prime}=n_{\nu;\mu}\bar{m}^{\nu}m^{\mu},
\end{align}
where the symbol $(;)$ represents the covariant derivative, $l_{\mu}$ and $n_{\mu}$ are the covariant counterparts of the null vectors $l^{\mu}$ and $n^{\mu}$.
  Thus, after substituting the values, we get the spin coefficients as
\begin{align}
\rho&=\frac{1}{r}+\frac{ia\cos\theta}{r^{2}}-\frac{a^{2}\cos^{2}\theta}{r^{3}},
\label{46a}\\
\rho^{\prime}&=-\frac{\Big(r-2M\Big)}{2r^{2}}-\frac{ia\cos\theta\Big(r-2M\Big)}{2r^{3}}
+\frac{a^{2}}{2r^{4}}\Bigg[2r\cos^{2}\theta-4M\cos^{2}\theta-r\Bigg].
\label{46b}
\end{align}
These spin coefficients are in the form of complex quantities. For $\mathbf{l}$ and $\mathbf{n}
$ to be hyper surface orthogonal, we require these spin coefficients to be real \cite{cha,grif}. For this purpose the coordinate system
needs to be rotated such that
\begin{equation}
m^{0}=0=m^{1}. \label{47}
\end{equation}
Then $\rho$ and $\acute{\rho}$ become real. Two rotations are performed for
this purpose.
First we do a type \textrm{II} and then a type \textrm{I} rotation \cite{cha}
leading to the new transformed tetrad as
\begin{align}
l^{\mu}&\rightarrow l^{\mu}+\bar{\beta}m^{\mu}+\beta\bar{m}^{\mu}
+\beta\bar{\beta}n^{\mu},\label{50}\\
n^{\mu} &\rightarrow n^{\mu}+\bar{\alpha}\left(  m^{\mu}+\beta n^{\mu
}\right)+\alpha\left(\bar{m}^{\mu}+\bar{\beta}n^{\mu}\right)
+\alpha
\bar{\alpha}\left(l^{\mu}+\bar{\beta}m^{\mu}+\beta\bar{m}^{\mu}+\beta
\bar{\beta}n^{\mu}\right),
\label{52}\\
m^{\mu}&\rightarrow m^{\mu}(1+\alpha\bar{\beta})+\alpha\beta\bar{m}^{\mu
}+n^{\mu}(\beta+\alpha\beta\bar{\beta})+\alpha l^{\mu}.\label{51}
\end{align}
where $\alpha$ and $\beta$ are complex functions to be determined such that Eq. (\ref{47}) is satisfied.
The spin coefficients
$\rho$ and $\rho^{\prime}$ as determined from the tetrad given in Eqs. (\ref{50})-(\ref{51}) are
\begin{align}
\rho&=\frac{1}{r}+\frac{a^{2}}{4r^{4}}\Bigg[-r\cos^{2}\theta-3r-4M+4M\cos
^{2}\theta\Bigg],
\label{62}\\
\rho^{\prime}&=-\frac{1}{2}\frac{\Big(r-2M\Big)}{r^{2}}+\frac{a^{2}\Big(r-4M\Big)}{2r^{4}
}-\frac{a^{2}\sin^{2}\theta}{8r^{5}}
\Big(7r^{2}+16M^{2}-22Mr\Big).
\label{63a}
\end{align}
As mentioned earlier, we are considering the surface defined by $r=$ constant and $t=$ constant, and the induced metric for the surface in this case has the components given by $g_{\theta\theta}$ and $g_{\phi\phi}$ of the metric (\ref{8}). The surface area element is
\begin{equation}
dS=\sqrt{g_{\theta\theta}g_{\phi\phi}}d\theta d\phi
=\sin
\theta\sqrt{\Sigma\Big(r^{2}+a^{2}+\frac{2a^{2}M\sin^{2}\theta}{r}\Big)}d\theta d\phi.
\label{64}
\end{equation}
The surface area is $A=\int_{0}^{\pi}\int_{0}^{2\pi}dS.$ From Eqs. (\ref{62})-(\ref{64}) it is clear that $\rho,\rho^{\prime}$ and the area element are independent of the CS coupling constant $\gamma$. As a result, the Hawking mass is independent of  $\gamma$ and is given  by
\begin{equation}
m_{H}=M-\frac{M^{2}a^{2}}{r^{3}}. \label{65}
\end{equation}
This shows the dependence of Hawking mass on $M,a$ and
 $r$ up to order $a^2$. This result matches with the Hawking mass for the Kerr metric \cite{Msc}.
It also shows that Hawking
mass is independent of the CS coupling constant $\gamma$. Here it is important to mention that up to the order $a\gamma^2$, the scalar field $\varphi$ does not contribute to the total energy of the spacetime \cite{12} (it does contribute to total energy of the spacetime, but the contribution is of order $a^2\gamma^2$ which is beyond the order considered here). This is also evident from Eq. (\ref{65}) which shows the $\gamma$ independent behaviour.

\section{The centre of mass energy}\label{ECM}
This section deals with the centre of mass energy $E_{CM}$ for the
collision of two neutral particles with equal masses i.e. $m_{1}=m_{2}=m_{0}$
in the vicinity of a slowly rotating Chern-Simons black hole. The particles
are moving from infinity with equal energies $E_{1}/m_{1}=E_{2}/m_{2}=1$
towards the black hole with different angular momenta $L_{1}$ and $L_{2}.$ The
motion as well as collision of the particles takes place in the equatorial
plane ($\theta=\pi/2)$. The expression for the $E_{CM}$  given by Ba\~{n}ados, Silk and West (BSW) \cite{20a} is
\begin{equation}
\frac{E_{CM}^{2}}{2m_{0}^{2}}=1-g_{\mu\nu}u_{1}^{\mu}u_{2}^{\nu}, \label{70}
\end{equation}
where $u_{1}^{\mu}$ =($\dot{t}_{1},\dot{r}_{1},\dot{\theta}_{1},\dot{\phi}
_{1})$ and $u_{2}^{\mu}=$($\dot{t}_{2},\dot{r}_{2},\dot{\theta}_{2},\dot{\phi
}_{2})$ represent the 4-velocity of the first and second particles. Here the overdot represents the derivative w. r. t the proper time $\tau$.
This formula is valid both for curved and flat spacetimes. For motion in the
equatorial plane $\dot{\theta}_{1}=\dot{\theta}_{2}=0.$ By Giving variation
0-3 to indices $\mu$ and $\nu$ in Eq. (\ref{70}), one obtains
\begin{equation}
\frac{E_{CM}^{2}}{2m_{0}^{2}}=1-\Big(g_{tt}\dot{t}_{1}+g_{t\phi}\dot{\phi}
_{1}\Big)\dot{t}_{2}-g_{rr}\dot{r}_{1}\dot{r}_{2}-\dot{\phi}_{2}\Big(g_{t\phi}\dot{t}_{1}
+g_{\phi\phi}\dot{\phi}_{1}\Big). \label{70A}
\end{equation}
The time-like geodesics for a particle of mass $m$ are \cite{12a,19a}
\begin{align}
\frac{dt}{d\tau}&=\frac{r\varepsilon}{r-2M}-\frac{2a\pounds M}
{r^{2}\Big(r-2M\Big)}+\frac{10a\pi\pounds \gamma^{2}}{r^{5}\Big(r-2M\Big)}\left[1+\frac{12M}
{7r}+\frac{27M^{2}}{10r^{2}}\right]-\frac{4\varepsilon M^{2}a^{2}}{r^{2}\Big(r-2M\Big)^{2}
},\label{cm1}\\
\frac{d\phi}{d\tau} &=\frac{\pounds }{r^{2}}+\frac{2a\varepsilon M}
{r^{2}\Big(r-2M\Big)}-\frac{10a\pi\varepsilon\gamma^{2}}{r^{5}\Big(r-2M\Big)}\Bigg[1+\frac{12M}
{7r}+\frac{27M^{2}}{10r^{2}}\Bigg]-\frac{\pounds a^{2}}{r^{3}\Big(r-2M\Big)},
\label{cm2}\\
\Big(\frac{dr}{d\tau}\Big)^{2}&=\varepsilon^{2}+\frac{\Big(2M-r\Big)\Big(\pounds ^{2}+r^{2}
\Big)}{r^{3}}-\frac{4aM\varepsilon\pounds }{r^{3}}+\frac{20a\pi \gamma^{2}}{r^{6}} \varepsilon
\pounds\Bigg[1+\frac{12M}{7r}+\frac{27M^{2}}{10r^{2}}
\Bigg]\nonumber\\&+a^{2}\Big(\frac{-r+\varepsilon^{2}\Big(r+2M\Big)}{r^{3}}\Big), \label{cm3}
\end{align}
where $\varepsilon=E/m$, $\pounds =L/m,E$ is the
energy and $L$ is the angular momentum of the particle. These equations are
velocity components of a particle of mass $m$. After substituting values
of the components of the 4-velocities of particles from Eqs.
(\ref{cm1})-(\ref{cm3}), we obtain the expression for the $E_{CM}^{2}$ as
\begin{align}
E_{CM}^{2} &{=}2m_{0}^{2}\Bigg[1{-}\frac{\pounds _{1}\pounds
_{2}}{r^{2}
}{+}\frac{r}{r{-}2M}{-}\frac{\sqrt{S_{1}S_{2}}}{r^{2}\Big(r{-}2M\Big)
}{+}a\Bigg[{-}14Mr^{5}{+}189M^{2}\gamma^{2}\pi{+}120Mr\gamma^{2}\pi
{+}70\pi r^{2}\gamma
^{2}\Bigg]\nonumber\\
& \times \Bigg[\frac{\pounds _{1}{+}\pounds
_{2}}{7r^{7}\Big(r-2M\Big)}-\frac{2\pounds _{1}^{2}\pounds
_{2}M+2\pounds _{2}^{2}\pounds _{1} M-\pounds _{1}^{2}\pounds
_{2}r-\pounds _{2}^{2}\pounds _{1}r+2\pounds _{1} r^{2}M+2\pounds
_{2}r^{2}M}{7r^{7}\Big(r-2M\Big)\sqrt{S_{1}S_{2}}
}\Bigg]\nonumber\\&+\frac{a^{2}}{2r^{4}\Big(r-2M\Big)^{2}}\Bigg\{-8M^{2}r^{2}+2\pounds
_{1} \pounds _{2} \Big(r^{2}-2Mr\Big)-\frac{8\pounds _{1}\pounds
_{2}M^{2}r^{2}\Big(r-2M\Big)}
{\sqrt{S_{1}S_{2}}}\nonumber\\&+r\sqrt{S_{1}S_{2}}\Big[\frac{2\pounds
_{1}^{2}Mr^{2}\Big( 2M-r\Big)+8M^{3}r^{3}+\pounds
_{1}^{4}\Big(r-2M\Big)  ^{2}}{S_{1} ^{2}}\nonumber\\&+\frac{2\pounds
_{2}^{2}Mr^{2}\Big(2M-r\Big)+8M^{3}r^{3} +\pounds
_{2}^{4}\Big(r-2M\Big)^{2}}{S_{2}^{2}}\Big]\Bigg\}\Bigg]
,\label{ECK}
\end{align}
where $\pounds _{1}$ and $\pounds _{2}$ are the angular momenta of
the particles, and $S_{1}=$ $2Mr^{2}-\pounds
_{1}^{2}\left(r-2M\right)$,\\$S_{2}=$ $2Mr^{2}-\pounds
_{2}^{2}\left(r-2M\right). $
Setting $\gamma=0$, one gets the $E_{CM}$ for
two particles of equal masses at the equatorial plane up to second
order in the spin parameter \cite{20a}.
The $a\rightarrow 0$ limit
recovers the Schwarzschild metric, so does the case $a=0=\gamma$.
The centre of mass energy in such a situation is \cite{20a}
\begin{equation}
E^2_{CM}=2m_{0}^{2}\Big[\frac{2r^2(r-M)-(r-2M)\pounds _{1}\pounds
_{2}-\sqrt{S_{1}S_{2}}}{r^2(r-2M)}\Big].
\end{equation}
 Eq. (\ref{ECK}) shows the dependence of $E_{CM}$ on $a$
and $\gamma^{2}.$ The $r\rightarrow\infty$ limit of Eq. (\ref{ECK}) gives $E_{CM}=2m_{0},$ which is the same as if the
particles are colliding in a flat spacetime. The event horizon of
the slowly rotating CS black hole is at $r_{H}=r_{H(Kerr)}$  where
$r_{H(Kerr)}$ is the event horizon of the Kerr metric \cite{12}. To
the required order in the spin parameter, the event horizon can be
written as $r_{H}\simeq2M-a^{2}/2M.$ From Eq. (\ref{ECK}), we see
that $E_{CM}$ is finite at $r=$ $r_{H}$ for finite
$\pounds _{1}$ and $\pounds _{2}$ in this slow rotation limit. As
the expression shows that it might diverge at $r=2M$, we take limit
$r\rightarrow2M,$ and get the expression
\begin{align}
&E_{CM}(r\rightarrow2M){=}\nonumber\\&\frac{m_{0}}{32\sqrt{7}M^{4}}\Bigg[1792M^{6}
\Big[\Big(\pounds _{1}{-}\pounds _{2}\Big)^{2}{+}16M^{2}\Big]{+}a\Big(\pounds _{1}{-}\pounds _{2}
\Big)^{2}\Big(\pounds _{1}
{+}\pounds _{2}\Big)\Big(448 M^{4}{-}709\pi\gamma^{2}\Big)\nonumber
\\
&
{+}28a^{2}M^{2}
\Big(5\pounds _{1}^{2}
{+}6\pounds _{1}\pounds _{2}{+}5\pounds _{2}^{2}{-}16M^{2}\Big)
\Big(\pounds _{1}{-}\pounds _{2}\Big)^{2}\Bigg]^{1/2},
\end{align}
which is also finite for finite values of $\pounds _{1}$ and $\pounds _{2}.$
\begin{figure}[!hptb]
\centering
\includegraphics[scale=0.8]{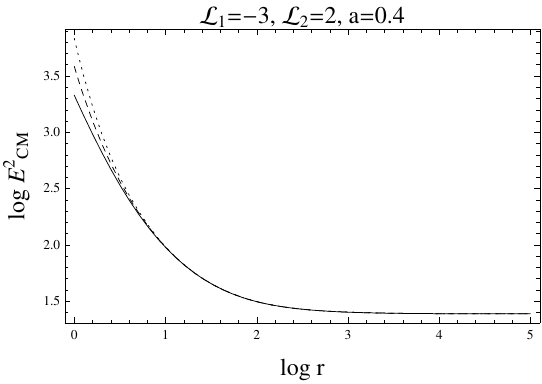}
\includegraphics[scale=0.8]{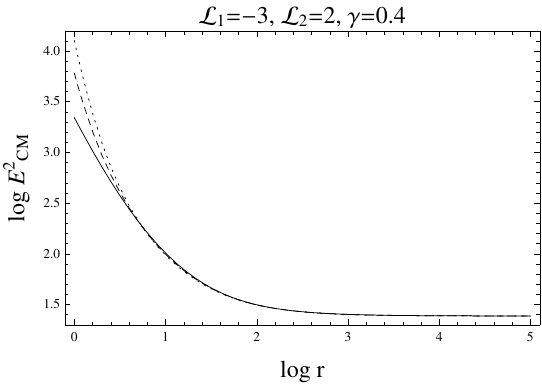}
\caption{Radial plot of $E_{CM}$ for some values of $\gamma,a,\pounds _{1},\pounds _{2} $.
On the left panel, the dotted, dashed and solid curves correspond to the values $0.3,0.2,0,$ respectively of the CS coupling constant $\gamma$. On the right panel, dotted, dashed and solid curves represent the centre of mass energy corresponding to the values $0.4,0.2,0,$ respectively of the spin parameter $a$.}
\end{figure}
The profiles of $E_{CM}$ have been plotted in Figure 1. From the graphs it is clear that the $E_{CM}$ increases with increase
in the coupling constant $\gamma$ and rotation parameter $a.$ Variation
between different curves is obvious for small values of radius but as $r$ increases, all curves merge together.
\section{Extraction of energy from slowly rotating Chern-Simons black hole
through Penrose process}\label{PP}

Penrose \cite{penrose} proposed a mechanism for extraction of energy from a
rotating black hole. It is based on the existence of negative energy orbits in
the ergosphere. Consider a positive energy particle moving along a time-like
geodesic into the ergosphere. The particle decays into two photons, one
crossing the event horizon and the other escaping to infinity. The photon
crossing the event horizon carries negative energy while the other one carries
more positive energy than the initial particle. It is assumed that such a
decay occurs at the turning point of the equatorial radial geodesics where
$\dot{r}=0$, then we have from Eq. (\ref{cm3})
\begin{equation}
\left[r\Big(r^{2}{+}a^{2}\Big){+}2Ma^{2}\right]E^{2}{-}2aLE\Big\lbrace2M{-}\frac{10\pi\gamma^{2}}{r^{3}
}
\Big[1{+}\frac{12M}{7r}{+}\frac{27M^{2}}{10r^{2}}\Big]\Big\rbrace{-}L^{2}\Big(r{-}2M\Big){-}m^{2}r\Delta{=}0.
\label{p1}
\end{equation}
The above equation is quadratic both in $E$
and $L$, so we solve for both. The solution in terms of $E$ leads to
\begin{align}
E&=\frac{1}{[r(r^{2}+a^{2})+2Ma^{2}]}\Bigg\lbrace L\Big[2aM-\frac{10a\pi\gamma^{2}}{r^{3}}
\Big(1+\frac{12M}{7r}
+\frac{27M^{2}}{10r^{2}}\Big)\Big]\pm\Big[L^{2}z+L^{2}r^{2}\Delta
\nonumber
\\
&
+m^{2}\Delta r(r(r^{2}
+a^{2})+2Ma^{2})\Big]^\frac{1}{2}\Bigg\rbrace,
\label{p2}
\end{align}
and that for the angular momentum is given by
\begin{equation}
L=
\frac{1}{r-2M}\Bigg\lbrace{-}E\Big[2aM{-}\frac{10a\pi\gamma^{2}}{r^{3}}\Big(1{+}\frac{12M}{7r}
{+}\frac{27M^{2}}{10r^{2}}\Big)\Big]{\pm}
\sqrt{E^{2}z+E^{2}r^{2}\Delta-m^{2}
\Delta(r^{2}-2Mr)}\Bigg\rbrace,
\label{p3}
\end{equation}
where
\begin{equation}
z=\frac{100a^2\pi^2\gamma^{4}}{r^6}\Bigg[1+\frac{12M}{7r}
+\frac{27M^{2}}{10r^{2}}\Bigg]^2 -\frac{40a^2M\pi\gamma^{2}}{r^3}\Bigg[1+\frac{12M}{7r}
+\frac{27M^{2}}{10r^{2}}\Bigg],
\end{equation}
 and the following identity was used for simplification
\begin{equation}
r^{2}\Delta-4Ma^{2}=\Big[r^{2}(r^{2}+a^{2})+2Ma^{2}r\Big]\Big(1-\frac{2M}{r}\Big). \label{p4}
\end{equation}
Only positive sign is selected in Eq. (\ref{p2}) because we want the
4-momentum of the particle to be future directed.
The orbit of the particle with negative energy in the ergosphere region is important for energy extraction in the Penrose process. From Eq. (\ref{p2}), for negative energy, we have the following conditions
\begin{equation}
L<0,\label{en1}
\end{equation}
\begin{equation}
\Big[r^{2}(r^{2}+a^{2})+2Ma^{2}r\Big]\Big\lbrace L^{2}(1-\frac{2M}{r})+m^{2}\Delta\Big\rbrace<0.\label{en2}
\end{equation}
These imply that conditions for negative energy are
\begin{equation}
E<0\Longleftrightarrow L<0,\label{en3}
\end{equation}
\begin{equation}
r-2M<-\frac{m^{2}\Delta r}{L^{2}}.\label{en4}
\end{equation}
Same conditions on energy have been observed for the Kerr black hole also. The terms containing the CS coupling parameter do not appear here as they get cancelled out in the process of simplification.

\begin{figure}[!hptb]
\centering
\includegraphics[scale=0.8]{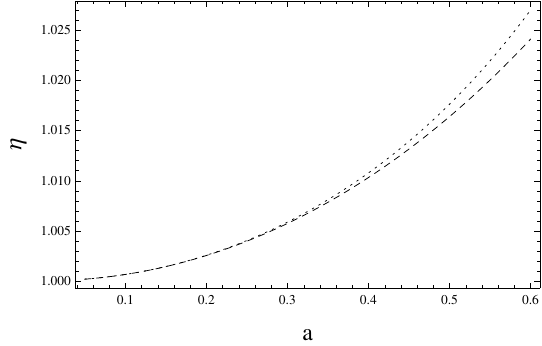}
\caption{Graph showing the efficiency of the Penrose process for the Kerr metric and the metric in slow rotation approximation, plotted against spin parameter $a$. The dotted curve shows the efficiency in Kerr's case and dashed curve is for approximated metric. Here, $M=1$. }
\label{pen}
\end{figure}

 As mentioned earlier we consider the situation where a particle of mass $m$
decays into two photons, one of which goes inside the event horizon and the other
one escapes to infinity. The photon which crosses the event horizon has
negative energy and the other photon carries more energy than the initial
particle. Let $E^{(0)},E^{(1)}$ and $E^{(2)}$ denote the energies of the
initial particle and photons respectively and $L^{(0)},L^{(1)}~$\ and
$L^{(2)}$ are their angular momenta. Let us take $m=1=E^{(0)}$ and $m=0$ for the initial particle and photons respectively. The
angular momentum of these particles can be obtained from Eq. (\ref{p3})
\begin{equation}
L^{(0)}=\frac{1}{r-2M}\Bigg\lbrace-2aM+\frac{10a\pi\gamma^{2}}{r^{3}}\Big[1+\frac{12M}
{7r}+\frac{27M^{2}}{10r^{2}}\Big]+ \sqrt{z+2Mr\Delta}\Bigg\rbrace
=
\alpha^{(0)}
,
\end{equation}
\begin{equation}
L^{(1)}
=
-\frac{1}{r-2M}\Bigg\lbrace2aM-\frac{10a\pi\gamma^{2}}{r^{3}}\Big[1+\frac
{12M}{7r}+\frac{27M^{2}}{10r^{2}}\Big]+\sqrt{z+r^2\Delta}\Bigg\rbrace E^{(1)}=\alpha^{(1)}
E^{(1)}
,
\end{equation}
\begin{equation}
L^{(2)} =-\frac{1}{r-2M}\Bigg\lbrace2aM-\frac{10a\pi\gamma^{2}}{r^{3}}\Big[1+\frac
{12M}{7r}+\frac{27M^{2}}{10r^{2}}\Big]-\sqrt{z+r^2\Delta}\Bigg\rbrace E^{(2)}=\alpha^{(2)}E^{(2)}
\ .
\end{equation}
According to the law of conservation of energy and angular momentum
\begin{equation}
E^{(0)}=E^{(1)}+E^{(2)},
\end{equation}
and
\begin{equation}
L^{(0)}=L^{(1)}+L^{(2)}=\alpha^{(0)}=\alpha^{(1)}E^{(1)}+\alpha^{(2)}E^{(2)}.
\end{equation}
Solving the above system of equations for $E^{(1)}$ and $E^{(2)}$ gives
\begin{align}
E^{(1)} & =-\frac{1}{2}\Bigg[\sqrt{\frac{z+2Mr\Delta}{z+r^2\Delta}}-1\Bigg]
\\&\simeq-\frac{1}{2}\Bigg[\sqrt{\frac{2M}{r}}-1\Bigg]+O(a^2\gamma^2),
\end{align}
\begin{align}
E^{(2)} & =\frac{1}{2}\Bigg[\sqrt{\frac{z+2Mr\Delta}{z+r^2\Delta}}+1\Bigg]
\\&\simeq\frac{1}{2}\Bigg[\sqrt{\frac{2M}{r}}+1\Bigg]+O(a^2\gamma^2).
\end{align}
The approximation in these energy values is done so that we can have their expressions up to our order of interest (up to $O(a^2)$ in spin parameter and $O(a \gamma^2)$ in the coupling parameter). Keeping the terms up to these orders, $E^{(1)}$ and $E^{(2)}$ are the same as in Kerr metric.
The gain in energy $\Delta E=\frac{1}
{2}[\sqrt{2M/r}-1]=-E^{(1)}.$ The efficiency of the energy extraction
by the Penrose process is given by
\[
\eta=\frac{E^{(0)}+\Delta E}{E^{(0)}}=\frac{1}{2}\Big(1+\sqrt{\frac{2M}{r}}\Big).
\]
 For maximum efficiency one must consider the situation where the radial distance
is minimum. Therefore we consider the situation where $r=r_{H}.$
For Kerr metric the maximum efficiency is found to be $\eta_{Kerr}=1.207$ (which corresponds to  $a=M$) \cite{cha}.
For slow rotation approximation, we cannot have $a=M$, therefore for metric (\ref8) the maximum efficiency is less than $\eta_{Kerr}$ and is independent of the CS coupling constant $\gamma$ as shown in Figure. \ref{pen}.
\section{Rotation of polarization vector for slowly rotating Chern-Simons black hole}\label{rotpol}
In this section our focus is the geometrical optics. The shadow and gravitational lensing of metric (\ref{8}) has been
studied in Refs. \cite{null} and \cite{lens} respectively. In this paper we study the
effect of polarization vector on the slowly rotating black hole in the
dynamical CS modified gravity. The study for the polarization
vector has been done initially for the Kerr metric in the small
rotation approximation $a<<M$ \cite{21}. This has been extended for rotating black
holes in a Randall-Sundrum brane and non-Kerr black holes \cite{23,24}. The formulation which is being used  here has been developed in Ref. \cite{21}, where
Newman-Penrose formalism is employed to determine optical quantities in the weak-field and slow rotation approximation.
The relationship between the tangent vector
$l^{\mu}$ (the wave vector) to the null congruence and the polarization vector
$f^{\mu}$ is given by the equations
\[
l^{\mu}l_{\mu}=0,\text{ \ \ }Dl^{\mu}=0,
\]
and
\[
l^{\mu}f_{\mu}=0,\text{ \ \ }Df^{\mu}=0.
\]
Here the operator $D$ is the covariant derivative in the direction of the vector $l^{\mu}$ .

The Newman-Penrose formalism assumes a null tetrad given by
 ${e_{a\mu}}=\{l_{\mu},n_{\mu}, m_{\mu},\bar{m}_{\mu}\}$ where
 \begin{equation}m^{\mu}=\frac{1}{\sqrt{2}}(a^{\mu}+ib^{\mu}).\end {equation}
The vector $m_{\mu}$ is important for the determination of polarization
vector. Consider the null rotations
\[
l^{\prime\mu}=Al^{\mu},\quad m^{\prime\mu}=e^{-i\chi}(m^{\mu}+Bl^{\mu}), \quad n^{\prime\mu}=\frac{1}{A}(n^{\mu}+B\bar{m}^{\mu}+\bar{B}m^{\mu}+B\bar{B}l^{\mu}),
\]
where $A>0,B$ is complex and $\chi$ is real.

\begin{figure}[!hptb]
\centering
\includegraphics[scale=0.8]{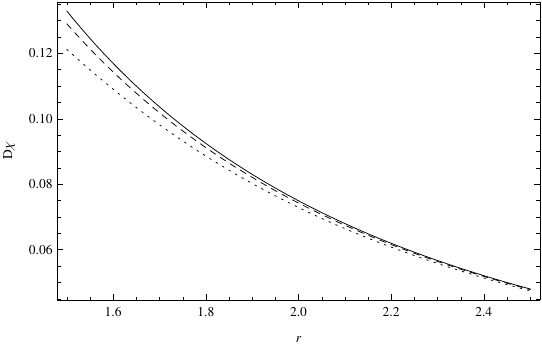}
\caption{Graph showing radial dependence of the polarization vector for various values of the CS coupling constant in the equatorial plane. For the solid curve $\gamma=0.01$, for dashed curve
$\gamma=0.03$ and for dotted curve $\gamma=0.05$. Here, $M=1, a=0.1$. }
\label{polar1}
\end{figure}

\begin{figure}[!hptb]
\centering
\includegraphics[scale=1.4]{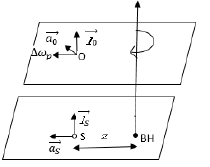}
\caption{The distance of the source $S$ and the observer $O$ from the
symmetry axis is denoted by $z$. Here $\vec{l}$ and $\vec{a}$ are tangent vectors to null congruence
and polarization vector respectively. (BH denotes black hole.) The variation of the polarization vector (Eq. (\ref{b}) is represented in this case by $\Delta \omega_{p}$. }
\label{polar2}
\end{figure}

If $l^{\mu}$ is tangent to the null congruence then the spin
coefficient $k=-Dl_{\mu}m^{\mu}$ becomes zero \cite{20}. For the parallel propagation of
the null tetrad along the null congruence, this condition leads to the vanishing
of the spin coefficients $\varepsilon$ and $\pi.$ Then one can identify the
plane generated by $l^{\mu}$ and $a^{\mu}$ with the polarization plane which
propagates along the $l^{\mu}$ direction. In other words, the polarization
vector can be identified with the $a^{\mu}$ vector leading to the construction of an orthonormal frame
\[
\{e_{a}^{(\mu)}\}=\{r^{(\mu)},\bar{r}^{(\mu)},q^{(\mu)},p^{(\mu)}\},
\]
such that this tetrad corresponds to the one forms $\omega^{(0)}=e^{\nu
}dt,\omega^{(1)}=e^{\lambda}dr,
\omega^{(2)}=e^{\mu}d\theta, \omega
^{(3)}=e^{\psi}(d\phi-\Omega dt),$ of the locally non-rotating frame (LNFR)
\cite{25}. The LNRF indices are written in parentheses. Consider the situation
where the source and the observer are at rest in LNFR, then they are rotating
with the black hole. 
The
vector $m_{+}^{(\mu)}$ (i.e. $a^{\mu})$ -- the projection of the vector $m^{\mu
}$ on LNRF is \cite{21}
\begin{align*}
a_{+}^{(\mu)} &  =\frac{1}{\sqrt{2}}\Big(0,-\frac{l^{(2)}}{l^{(0)}},1-W(l^{(2)}%
)^{2},Wl^{(2)}l^{(3)}\Big),\\
b_{+}^{(\mu)} &  =\frac{1}{\sqrt{2}}\Big(0,-\frac{l^{(3)}}{l^{(0)}},-Wl^{(2)}%
l^{(3)},1-W(l^{(3)})^{2}\Big),
\end{align*}
where $W=1/[l^{(0)}(l^{(0)}+l^{(1)})]$ and $l^{(\mu)}$ represents the projection
of $l^{\mu}$ on LNRF. A null rotation of the form
\begin{equation}
m_{+}^{(\mu)}\rightarrow m^{(\mu)}=e^{-i\chi}m_{+}^{(\mu)},\label{a1}%
\end{equation}
is done to make the spin coefficient $\varepsilon$ zero. The expression for
$\varepsilon=Dm_{\mu}\bar{m}^{\mu}/2$ and the rotation mentioned above
lead to the expression
\[
D\chi=-2i\varepsilon_{+}.
\]
This equation gives the variation of angle $\chi$ in the direction of
$l^{\mu}.$ For the metric (\ref{8})  $D\chi$ is
\begin{align}
D\chi &  =-2i\varepsilon_{+}=\frac{\Big(\Gamma_{(\theta)(t)}^{(t)}l^{(t)}
+\Gamma_{(\theta)(r)}^{(r)}l^{(r)}+\Gamma_{(\theta)(\theta)}^{(r)}l^{(\theta
)}+\Gamma_{(\theta)(\varphi)}^{(t)}l^{(\varphi)}\Big)l^{(\varphi)}}{l^{(t)}
+l^{(r)}}\nonumber\\
&  -\frac{\Big(\Gamma_{(\varphi)(t)}^{(r)}l^{(t)}+\Gamma_{(\varphi)(r)}
^{(t)}l^{(r)}+\Gamma_{(\varphi)(\theta)}^{(t)}l^{(\theta)}+\Gamma_{(\varphi
)(\varphi)}^{(r)}l^{(\varphi)}\Big)l^{(\theta)}}{l^{(t)}+l^{(r)}}+\Gamma
_{(\varphi)(\varphi)}^{(\theta)}l^{(\varphi)}+\Gamma_{(\varphi)(t)}^{(\theta
)}l^{(t)}.\label{a22}
\end{align}
After substituting the values of $\Gamma_{(\nu)(\delta)}^{(\mu)}$ and $l^{\mu}$ provided in the Appendix, $D\chi$ up to the linear order in the spin parameter takes the form
\[
D\chi=\frac{-L\cos\theta}{r^{2}\sin^{2}\theta}+\Big(\frac{3Mr^{5}-6\pi\gamma
^{2}(18M^{2}+10Mr+5r^{2})}{r^{9}}\Big)a\sin\theta\sqrt{Q-L^{2}\cot^{2}\theta}.\label{polar}
\]
Up to the order $a$, $d\theta/d\lambda$ reduces to \cite{null}
\begin{equation}
\frac{d\theta}{d\lambda}=\frac {\sqrt{Q-L^{2}\cot^{2}\theta}}{r^2}.
\end{equation}
Thus $D\chi$ modifies to
\begin{equation}
D\chi=\frac{-L\cos\theta}{r^{2}\sin^{2}\theta}+\Big(3M-\frac{6\pi\gamma
^{2}(18M^{2}+10Mr+5r^{2})}{r^{5}}\Big)\frac{a\sin\theta}{r^{2}}\frac{d\theta
}{d\lambda}.\label{a}%
\end{equation}
Equation $\left(  \ref{a}\right)  $ is associated with the variation between
$a^{(\mu)}$ and $a_{+}^{(\mu)},$ according to the null rotation given in
Eq. (\ref{a1}).
Figure. \ref{polar1} shows $D\chi$ for some values of $\gamma$. From the graph it is evident that the polarization angle decreases as $\gamma$ increases.
Eq. (\ref{a}) can
be written as
\begin{equation}
D\chi=\frac{-L\cos\theta}{r^{2}\sin^{2}\theta}-\Big(3M-\frac{6\pi\gamma
^{2}(18M^{2}+10Mr+5r^{2})}{r^{5}}\Big)\frac{a}{r^{2}}\frac{d\cos\theta}{d\lambda
}.\label{d}%
\end{equation}
The total change in the polarization vector is
\begin{equation}
\Delta \omega=\Delta \chi+\Delta \varphi.\label{b}%
\end{equation}
Here, the second term comes from the spacetime dragging. The integration of Eq. (\ref{a}) with the help of the $\psi$ coordinate
gives $\Delta \chi.$ The $\psi$ coordinate  is taken as the
azimuthal angle in the orbital plane of the null coordinate \cite{21}. The angle between the equatorial and orbital planes is represented by $\alpha$.
 We have the expression of the form
\begin{equation}
\cos\theta=\sin\psi\sin\alpha.\label{c}
\end{equation}
Using Eq. (\ref{c}) in Eq. (\ref{d}), we have
\begin{equation}
D\chi=\frac{-L\cos\theta}{r^{2}\sin^{2}\theta}+D\chi^{\prime},
\end{equation}
where
\begin{equation}
D\chi^{\prime}=-\Big(3M-\frac{6\pi\gamma^{2}(18M^{2}+10Mr+5r^{2})}{r^{5}}
\Big)\frac{a\sin\alpha}{r^{2}}\frac{d\sin\psi}{d\lambda}.\label{e}
\end{equation}
Consider the physical scenario where the source is on the equatorial plane
($\theta=\alpha=\pi/2)$ and the observer is on the plane with $\theta<\pi/2.$
The distance of the observer and the source from the symmetry
axis is denoted by $z$.
This situation is drawn in the Figure. \ref{polar2}. The source and the observer are represented by the timelike curves
$r_{s}$ and $r_{o}$ respectively. Also, $L=0$ in the photon equations of motion.
Following the assumptions, one has the expression
\[
\sin\psi=\frac{\sqrt{r^{2}-z^{2}}}{r},
\]
which has the derivative of the form
\begin{equation}
\frac{d\sin\psi}{d\mu}=\frac{z^{2}}{r^{2}\sqrt{r^{2}-z^{2}}}\frac{dr}
{d\lambda}.\label{f}
\end{equation}
Substituting Eq. (\ref{f}) in Eq. (\ref{e}), one has
\begin{equation}
D\chi^{\prime}=-\Big(3M-\frac{6\pi\gamma^{2}(18M^{2}+10Mr+5r^{2})}{r^{5}}\Big)\frac{a}{r^{4}}\frac{z^{2}}{\sqrt{r^{2}-z^{2}}}\frac{dr}{d\lambda}.
\end{equation}
Integration leads to
\[
\int_{0}^{\chi}d\chi^{\prime}=\int_{z}^{r_{o}}-\Big(3M-\frac{6\pi\gamma
^{2}(18M^{2}+10Mr+5r^{2})}{r^{5}}\Big)\frac{a}{r^{4}}\frac{z^{2}}{\sqrt
{r^{2}-z^{2}}}dr,
\]
where $D\chi^{\prime}=l^{1}d\chi^{\prime}/dr$ and $l^{1}=dr/d\lambda$ which leads to
$D\chi^{\prime}=d\chi^{\prime}/{d\lambda}$.
\begin{align*}
\Delta \chi &  =\frac{a}{448z^{7}r_{o}^{8}}\Bigg[6615\pi^{2}\gamma^{2}
r_{o}^{8}M^{2}+2100\pi^{2}\gamma^{2}z^{2}r_{o}^{8}+\sqrt{r_{o}^{2}-z^{2}
}\Big(12288z\pi\gamma^{2}r_{o}^{7}M+4200d^{3}\pi\gamma^{2}r_{o}^{6}\\
&  +7056\pi\gamma^{2}M^{2}z^{5}r_{o}^{2}-896Mz^{5}r_{o}^{7}+13230\pi\gamma
^{2}M^{2}dr_{o}^{6}+8820\pi\gamma^{2}M^{2}z^{3}r_{o}^{4}\\
&  +6048\pi\gamma^{2}M^{2}z^{7}+2800\pi\gamma^{2}z^{5}r_{o}^{4}+2240\pi
\gamma^{2}z^{7}r_{o}^{2}+4608\pi\gamma^{2}Mz^{5}r_{o}^{3}+3840\pi\gamma
^{2}Mz^{7}r_{o}\\
&  -448Mz^{7}r_{o}^{5}+6144\pi\gamma^{2}Mz^{3}r_{o}^{5}\Big)+
\arctan(\frac{z}{\sqrt{r_{o}^{2}-z^{2}}})\Big(-4200\pi\gamma^{2}z^{2}r_{o}^{8}-13230\pi
\gamma^{2}M^{2}r_{o}^{8}\Big)\Bigg].
\end{align*}
Setting $\gamma=0$ and $r_{o}\rightarrow\infty $ gives $\Delta \chi=-2aM/z^{2}$ which is same as for the Kerr spacetime \cite{21}.

\section{Summary and conclusion}\label{SC}

The Kerr metric is that unique solution of the Einstein field equations which obeys
the no-hair theorem i.e. it depends on the mass and spin for its complete
description. In recent years several alternative theories of gravity have been
constructed in which the spacetimes are different from Kerr,
 having parameters other than mass and spin. Putting all
deviations to zero, all the modified spacetimes become Kerr.
One of these modified gravity theories is the Chern-Simons gravity theory
having two independent formulations namely, the non-dynamical theory and
dynamical theory. Black hole solutions have been developed in both
formulations but our focus in this paper is the solution of the CS
gravity given in Ref. \cite{12}. The solution has been developed under the
assumptions of small rotation parameter and small coupling constant. Setting
the coupling constant equal to zero, the solution reduces to Kerr in small
rotation approximation. Therefore, our focus in this paper is to study the effects
of the coupling constant in  different physical situations. First we
studied the Hawking mass outside the event horizon of the black hole and obtained an exact value.
We find that the Hawking mass is independent of the CS coupling constant $\gamma.$ It
just depends on the mass and spin parameter like in Kerr's case. Next, based
on the BSW mechanism \cite{20a}, we studied the dependence of the coupling constant on
the $E_{CM}$ for two neutral colliding particles of equal masses.
The graphs are drawn for different values of the coupling constant and rotation parameter. The graphical results show that both $a$ and $\gamma$ cause an increase in the centre of mass energy. We also studied the energy extraction through Penrose process and found that the efficiency of the process is independent of the CS coupling constant $\gamma$ but is not exactly equal to the efficiency of the Kerr case due to the assumption of the small rotation approximation. In the geometrical optics regime, rotation of the polarization vector is studied. The polarization vector shows decreasing behaviour with increasing CS coupling constant.
\acknowledgments
RR acknowledges the support of the Indigenous Fellowship of the Higher Education Commission (HEC) of Pakistan. The work was also supported by the HEC, NRPU research grants  20-2087 and 6151.
\section{Appendix}
The components of $l^{\mu}=dx^{\mu}/d\lambda$ and its projections $l^{(\mu)}$ are related by
\begin{align}
&l^{(0)}=l^{(t)}=l^{0}e^{\nu}=e^{\nu}\dot{t},\nonumber \\&
l^{(1)}=l^{(r)}=e^{\lambda}l^{1}=e^{\lambda}\dot{r},
\nonumber \\&
l^{(2)}=l^{(\theta)}=e^{\mu}l^{2}=e^{\mu}\dot{\theta},
\nonumber \\&
l^{(3)}=l^{(\phi)}=e^{\psi}(l^{3}-\Omega l^{0})=e^{\psi}(\dot{\phi}-\Omega\dot{t}),
\nonumber \\&
\Omega   =\frac{2aM}{r^{3}}-\frac{10a\pi\gamma^{2}}{r^{6}}\Big(1+\frac{12}%
{7}\frac{M}{r}+\frac{27}{10}\frac{M^{2}}{r^{2}}\Big).
\end{align}

The functions $e^{\nu},%
 e^{\lambda}, %
 e^{\mu}$ and $e^{\psi }$ are
\begin{align*}
e^{2\nu} &  =1-\frac{2M}{r}+\frac{2a^{2}M}{r^{3}}\cos^{2}\theta+\frac
{4a^{2}M^{2}}{r^{4}}\sin^{2}\theta,\\
e^{2\lambda} &  =g_{rr}, \quad e^{2\mu}=g_{\theta\theta}, \quad e^{2\psi}=g_{\phi\phi}.
\end{align*}

The Christoffel symbols in LNFR are given as \cite{25}
\begin{align}
&\Gamma_{(r)(t)}^{(t)}=\Gamma_{(t)(t)}^{(r)}=e^{-\lambda}\partial_{r}
\nu,
\nonumber \\&
\Gamma_{(\theta)(t)}^{(t)}=\Gamma_{(t)(t)}^{(\theta)}=e^{-\mu}
\partial_{\theta}\nu,\nonumber \\&
\Gamma_{(\theta)(r)}^{(r)}=-\Gamma_{(r)(r)}^{(\theta)}=e^{-\mu}\partial_{\theta}\lambda,
\nonumber \\&
\Gamma_{(\theta)(\theta)}^{(r)}=-\Gamma_{(r)(\theta)}^{(\theta)}
=-e^{-\lambda}\partial_{r}\mu,\nonumber \\&
\Gamma_{(\phi)(\phi)}^{(r)}=-\Gamma_{(r)(\phi)}^{(\phi)}=-e^{-\lambda}\partial_{r}\psi,
\nonumber \\&
\Gamma_{(\phi)(\phi)}^{(\theta)}=-\Gamma_{(\theta)(\phi)}^{(\phi)}=-e^{-\mu}\partial_{\theta}\psi,
\nonumber\\&
 \Gamma_{(r)(\phi)}^{(t)}=\Gamma_{(t)(\phi)}^{(r)}=\Gamma_{(\phi)(r)}
^{(t)}=\Gamma_{(\phi)(t)}^{(r)}=-\Gamma_{(t)(r)}^{(\phi)}=-\Gamma
_{(r)(t)}^{(\phi)}=\frac{1}{2}e^{\psi-\nu-\lambda}\partial_{r}\Omega,
\nonumber\\&
\Gamma_{(\theta)(\phi)}^{(t)}=\Gamma_{(t)(\phi)}^{(\theta)}=\Gamma
_{(\phi)(\theta)}^{(t)}=\Gamma_{(\phi)(t)}^{(\theta)}=-\Gamma_{(t)(\theta
)}^{(\phi)}=-\Gamma_{(\theta)(t)}^{(\phi)}=\frac{1}{2}e^{\psi-\nu-\mu}%
\partial_{\theta}\Omega.
\end{align}

\end{document}